\documentclass[prl, aps, 12pt, showkeys, showpacs, tightenlines]{revtex4}
\usepackage{hyperref}
\begin{document}
\title{Proton Sea Quark Flavour Asymmetry and Roper Resonance}
\author{Yong-Jun Zhang}
\email{yong.j.zhang@gmail.com}
\affiliation{Science College, Liaoning Technical University, Fuxin, Liaoning 123000, China}

\author{Bin Zhang}
\email{zb@mail.tsinghua.edu.cn(Communication author)}
 \affiliation{ Department of Physics,
Tsinghua University, Beijing 100084, China} \affiliation{Center for
High Energy Physics, Tsinghua University, Beijing 100084, China}

\begin{abstract}
We study the proton and the Roper resonance together with the meson cloud model, by constructing a Hamiltonian matrix and solving the eigenvalue equation. The proton sea quark flavour asymmetry and some properties of the Roper resonance are thus reproduced in one scheme.
\end{abstract}
\keywords{proton, sea quark, flavour asymmetry, Roper, Hamiltonian matrix}
\pacs{14.20.Dh, 14.20.Gk, 14.65.Bt, 13.30.Eg}
\maketitle

\section{Introduction}
The proton flavour asymmetry has been observed by many experiments \cite{1,2,3,4}. One recent observation \cite{4} is $\bar{d}-\bar{u}=0.118\pm 0.012$. Thus, besides the three valance quarks, the proton contains sea quarks which have more $\bar{d}$ than $\bar{u}$. The meson cloud model \cite{meson1,meson2,meson3} is one of the many models that can explain the proton sea quark flavour asymmetry.

The Roper resonance, also known as N(1440), has the same quantum number as the nucleon and is a nucleon resonance. According to a simple quark model, it is a three-quark state that one quark is in a radical excited state. But then there is a parity reverse problem. So, N(1440) may be not just a $qqq$ state and may contain other components like $qqqq\bar{q}$. Juli\'{a}-D\'{i}az and Riska \cite{H} introduce $qqqq\bar{q}$ to both the nucleon and N(1440). By constructing a Hamiltonian matrix and calculating its eigenstates, they conclude that N(1440) has $qqqq\bar{q}$ admixture ranging from 3 to 25 \% depending on the constituent quark mass. 

In this letter, we study the proton and N(1440) together by using the meson cloud model and constructing a Hamiltonian matrix. 

\section{Meson Cloud Model with only bare nucleon and pions}
We start in this section with a simple meson cloud model that contains only bare nucleon and pions to show how to construct the Hamiltonian matrix and solve the corresponding eigenvalue equation.

According to the meson cloud model \cite{meson3}, the proton wavefunction is 
\begin{equation}\label{mcm}
        |p\rangle=C_1|p_0\rangle+C_2|N_0\pi_0\rangle+\cdots,
\end{equation}
in which each Fock state has the same isospin and parity as the proton's and so 
\begin{equation}
    |N_0\pi_0\rangle=-\sqrt{\frac{1}{3}}|p_0\pi_0^0\rangle+\sqrt{\frac{2}{3}}|n_0\pi_0^+\rangle.
\end{equation}
Here $\pi_0$ is bare pion; $p_0$, $n_0$ and $N_0$ are bare proton, bare neutron and bare nucleon. The bare particles are not exactly the same as the physical particles, they have the same quantum numbers but different energies. The bare particles have quark structure
\begin{equation}
\begin{array}{l}
    p_0\rightarrow uud,\\
    n_0\rightarrow udd,\\
    \pi_0^+\rightarrow u\bar{d},\\
    \pi_0^-\rightarrow \bar{u}d,\\
    \pi_0^0\rightarrow \frac{1}{2}(u\bar{u}+d\bar{d}).
\end{array}
\end{equation}

Because the proton sea quark flavour asymmetry mainly arises from the component $|N_0\pi_0\rangle$, by a guesswork one may write down the proton wavefunction as simple as
\begin{equation} \label{protonwavefunction}
        |p\rangle=0.91|p_0\rangle- 0.42|N_0\pi_0\rangle
\end{equation}
to reproduce the sea quark flavour asymmetry
\begin{equation} \label{flavour_asymmetry}
    \bar{d}-\bar{u}=0.118.
\end{equation}
One may also write down an orthogonal wavefunction
\begin{equation}\label{996} 
        |p^{\prime}\rangle=0.42|p_0\rangle+ 0.91|N_0\pi_0\rangle
\end{equation}
that $\langle p|p^{\prime}\rangle=0$. Given the proton state as the ground state, the orthogonal state can only be an excited state. Can it be N(1440)? Is its mass 1440MeV?
Then other Fock states like $|N_0\pi_0\pi_0\rangle$ would lead to more orthogonal states. All these states are associated with a common Hamiltonian matrix. 

We start with a simple Fock state basis containing only bare nucleon and pions
\begin{equation} 
\ \ \ \ \	\{\ \ |p_0\rangle,\ |N_0\pi_0\rangle,\ \ \ |N_0\pi_0\pi_0\rangle,\ \cdots\ \ \}.
\end{equation}
From this basis, we construct a Hamiltonian matrix
\begin{equation}\label{proton}
	\hat{H}=
        \left [
        \begin{array}{ccccccccc}
                E_{p_0}&         &h                    &		&0                      &\cdots  \\
                h      &       &E_{p_0}\!\!+\!E_{\pi_0}&         &h                    &\cdots  \\
                0      &         &h                    &		&E_{p_0}\!\!+\!2E_{\pi_0}      &\cdots  \\
                \vdots &         &\vdots               &  	&\vdots                 & \ddots\\
        \end{array}
        \right ]
\end{equation}
whose elements are parameterized as 
\begin{equation}
\begin{array}{cl}
    	\langle p_0|\hat{H}|p_0\rangle 				&=E_{p_0}, \\
    	\langle N_0\pi_0|\hat{H}|N_0\pi_0\rangle 		&= E_{p_0}+E_{\pi_0}, \\
    	\langle N_0\pi_0\pi_0|\hat{H}|N_0\pi_0\pi_0\rangle   	&\approx E_{p_0}+2E_{\pi_0}, \\
    	\cdots,\\
	\langle p_0|\hat{H}|N_0\pi_0\rangle			&= h,\\
	\langle N_0\pi_0|\hat{H}|N_0\pi_0\pi_0\rangle		&\approx h,\\
	\dots,\\
	\langle p_0|\hat{H}|N_0\pi_0\pi_0\rangle		&= 0,\\
	\langle p_0|\hat{H}|N_0\pi_0\pi_0\pi_0\rangle		&= 0,\\
	\langle p_0\pi_0|\hat{H}|N_0\pi_0\pi_0\pi_0\rangle	&= 0,\\
	\cdots.
\end{array}
\end{equation}
where $E_{p_0}$ is the energy of the bare proton, $E_{\pi_0}$ is the energy of the bare pion plus non-relative interaction effect between $N_0$ and $\pi_0$. 

Here only the single pion annihilation/creation interaction is considered. The parameter $h$ is about how often a bare pion is annihilated or created. In the limit $h\to 0$, there would be no bare pion created or annihilated, each Fock state along would become a physical state. For example, the Fock state $|N_0\pi_0\rangle$ would become a two-body state $N_0\pi_0$ whose energy would be $E_{p_0}\!+\!E_{\pi_0}$. When $h\neq 0$, all the Fock states start to mix together to form a new set of physical states which are actually the eigenstates of a Hamiltonian matrix. The matrix diagonal elements are the energies of each Fock state, for example, the element $\langle N_0\pi_0|\hat{H}|N_0\pi_0\rangle$ takes value $E_{p_0}\!+\!E_{\pi_0}$. The off-diagonal elements are controlled by $h$. Given the bare nucleon having the fixed energy $E_{p_0}$, $h$ is mainly affected by the number of $\pi_0$. In this letter, we only study the lowest few eigenstates which contains only few $\pi_0$ that are likely in different flavours and spacial states. Thus the pion exchange symmetry effect is neglected, and $h$ is treated as a constant.

The matrix (\ref{proton}) contains three parameters and we shall fix them by three observations, 
\begin{equation}\label{table1} 
        \begin{array}{l|cccccccccc}
	\hline
        M_{\rm proton}   		& 938\ {\rm MeV}    	  	\\
        M_{\rm N(1440)}    			& 1440\ {\rm MeV}            	\\
    	\bar{d}-\bar{u}                 & 0.118           		\\
	\hline
        \end{array}
\end{equation}
Before solving the corresponding eigenvalue equation, we need first to truncate the matrix from infinite size to a finite size $n\!\times\!n$ by using energy cut-off. With different $n$, the parameters are obtained as
\begin{equation} \label{table2}
        \begin{array}{c|cccccccccccccc}
        \hline 
	n&E_{p_0}&E_{\pi_0}&h\\
	\hline
		2   &    1027        &   324        & 192               \\
		3   &    1037        &   403        & 211               \\
		4   &    1038        &   412        & 215               \\
		5   &    1038        &   413        & 215               \\
	        \vdots&\vdots&\vdots&\vdots\\
	\hline
        \end{array}
\end{equation}
As $n$ increases, the parameters converge. So the energy cut-off is safe. With the fixed parameters
\begin{equation} \label{fixed3}
	E_{p_0}=1038{\rm MeV},\ E_{\pi_0}=413{\rm MeV},\ h=215{\rm MeV},	
\end{equation}
the wavefunctions are obtained as 
\begin{equation}\label{eigv1}
\begin{array}{c|cccccc}
\hline
E({\rm MeV})    & |p_0\rangle   &|N_0\pi_0\rangle   & |N_0\pi_0\pi_0\rangle &|N_0\pi_0\pi_0\pi_0\rangle  &\cdots   \\\hline
938	&	0.90        &	-0.42        &	0.10        &	-0.02           &	\cdots \\
1440	&	0.42        &	0.78        &	-0.46        &	0.12             &	\cdots \\
1863	&	0.12        &	0.45        &	0.75        &	-0.45           &	\cdots \\
\vdots          &\vdots     &\vdots         &\vdots   &\vdots                   &   \ddots     \\
\hline
\end{array}
\end{equation}
and their percentages of each Fock state are 
\begin{equation}\label{density_matrix1}
\begin{array}{c|cccccc}
\hline
E({\rm MeV})    & |p_0\rangle   &|N_0\pi_0\rangle   & |N_0\pi_0\pi_0\rangle &|N_0\pi_0\pi_0\pi_0\rangle  &\cdots   \\\hline
938	&	0.81        &	0.18        &	0.01        &	0.00        &	\cdots \\
1440	&	0.17        &	0.60        &	0.21        &	0.02        &	\cdots \\
1863	&	0.01        &	0.20        &	0.56        &	0.21        &	\cdots \\
\vdots          &\vdots     &\vdots         &\vdots   &\vdots                   &   \ddots     \\
\hline
\end{array}
\end{equation}
Or one explicitly  writes the wavefunctions of the proton and N(1440) as
\begin{equation} 
	\begin{array}{ll}
		|p\rangle		&=0.90|p_0\rangle-0.42|N_0\pi_0\rangle+0.10|N_0\pi_0\pi_0\rangle+\cdots,\\
		|{\rm N(1440)}\rangle	&=0.42|p_0\rangle+0.78|N_0\pi_0\rangle-0.46|N_0\pi_0\pi_0\rangle+\cdots,\\
	\end{array}
\end{equation}
and their percentages of each Fock state as
\begin{equation} 
	\begin{array}{ll}
		{\rm proton:}	&0.81|p_0\rangle\langle p_0|+0.18|N_0\pi_0\rangle\langle N_0\pi_0|+0.01|N_0\pi_0\pi_0\rangle\langle N_0\pi_0\pi_0|+\cdots,\\
		{\rm N(1440):}	&0.17|p_0\rangle\langle p_0|+0.60|N_0\pi_0\rangle\langle N_0\pi_0|+0.21|N_0\pi_0\pi_0\rangle\langle N_0\pi_0\pi_0|+\cdots.
	\end{array}
\end{equation}

\section{Meson Cloud Model With Bare $\Delta$}
In this section, we add in the Fock states that contain bare $\Delta$, or $\Delta_0$. Then the Fock state basis and the Hamiltonian matrix becomes
\begin{equation} 
        \{\ |p_0\rangle,\ \ \ |N_0\pi_0\rangle,\ \ \ |N_0\pi_0\pi_0\rangle,\ \ |N_0\pi_0\pi_0\pi_0\rangle,\ \cdots,  \ \ |\Delta_0\pi_0\rangle, \ \ |\Delta_0\pi_0\pi_0\rangle,\ \ \ \cdots \},
\end{equation}
\begin{equation}
        \hat{H}=
        \left [
        \begin{array}{ccccccccccccc}
                E_{p_0}         &h_1                    &0                      &0                      &\cdots &0             			&0                              &\cdots\\
                h_1             &E_{p_0}\!+\!E_{\pi_0}  &h_1                    &0                      &\cdots &0                      	&0                     		&\cdots\\
                0               &h_1                    &E_{p_0}\!+\!2E_{\pi_0} &h_1 			&\cdots &h_3                      	&0                              &\cdots\\
                0               &0                      &h_1 	 		&E_{p_0}\!+\!3E_{\pi_0} &\cdots &0                      	&h_3                            &\cdots\\
                \vdots          &\vdots                 &\vdots                 &\vdots                 &\ddots &\vdots                 	&\vdots                         &\vdots\\
                0      		&0                      &h_3                    &0                      &\cdots &E_{\Delta_0}\!+\!E_{\pi_0}     &h_2               		&\cdots\\
                0               &0             		&0                      &h_3                    &\cdots &h_2       			&E_{\Delta_0}\!+\!2E_{\pi_0} 	&\cdots\\
                \vdots          &\vdots                 &\vdots                 &\vdots                 &\ddots &\vdots                 	&\vdots                         &\ddots\\
        \end{array}
        \right ],
\end{equation}
the parameters are increased to be
\begin{equation} 
	\{\ E_{p_0},\ E_{\Delta_0},\ E_{\pi_0},\ h_1,\ h_2,\ h_3\ \}.
\end{equation}
The 6 parameters are to be fixed by 6 experimental observations\cite{PDG,4,Mohr08,Manley92,Cutkosky80,Hoehler79},
\begin{equation}\label{table3} 
        \begin{array}{l|c|c|cccc}
	\hline
							& {\rm Observations}	& {\rm To\ reproduce}(x_e)	& {\rm Uncertainty}(\sigma_x)	\\
	\hline
        M_{\rm proton}\ ({\rm MeV})   			& 938.272013\pm 0.000023    	  	& 938.272013					& 0.000023				\\
        M_{{\rm N(1440)}}   \ ({\rm MeV})			& 1420 {\rm \ to \ }1470            	& 1440					& 25				\\
        M_{{\rm N(1710)}}   \ ({\rm MeV}) 			& 1680 {\rm \ to \ }1740            	& 1710					& 30				\\
    	\bar{d}-\bar{u}                 		& 0.118\pm 0.012			& 0.118					& 0.012				\\
    	{\rm B}\left({\rm N(1440)}\to N\pi\right)         & 0.55-0.75           			& 0.65					& 0.10				\\
    	{\rm B}\left({\rm N(1440)}\to N\pi\pi\right)      & 30\!-\!40\%           		& 0.35					& 0.05				\\
	\hline
        \end{array}
\end{equation}
where the branching radios are to be estimated from the N(1440) wavefunction, $M_{\rm N(1710)}$ is to be reproduced as the third eigenvalue.
In the calculation, we fix the parameters by minimizing 
$    \chi^2=\sum \frac{(x-x_e)^2}{\sigma_x^2}$.

The parameters are obtained as
\begin{equation} \label{table22}
        \begin{array}{cc|cccccccccccccc}
        \hline 
	n_1 & n_2 &E_{p_0}&E_{\Delta_0}&E_{\pi_0}&h_1&h_2&h_3\\ 
	\hline
		4   &    2   &    1042        &   1405          & 425  & 221  & 156	& 148\\ 
		5   &    3   &    1042        &   1408          & 426  & 221  & 154	& 149\\ 
		6   &    4   &    1042        &   1408          & 426  & 221  & 154	& 149\\ 
	        \vdots&\vdots&\vdots&\vdots&\vdots&\vdots&\vdots&\vdots\\
	\hline
        \end{array}
\end{equation}
where $n_1$ is the number of $|N_0\cdots\rangle$ Fock states and $n_2$ is the number of $|\Delta_0\cdots\rangle$ Fock states. As $n_1$ and $n_2$ increase, the parameters converge. So the energy cut-off is safe. With the fixed parameters
\begin{equation}\label{fixed6} 
	E_{p_0}=1042,\ E_{\Delta_0}=1408,\ E_{\pi_0}=426,\ h_1=221,\ h_2=154,\ h_3=149,\	({\rm MeV})
\end{equation}
the wavefunctions are obtained as
\begin{equation}\label{eigv}
\begin{array}{c|ccccccccc}
\hline
E({\rm MeV})    & |p_0\rangle   &|N_0\pi_0\rangle   & |N_0\pi_0\pi_0\rangle &|N_0\pi_0\pi_0\pi_0\rangle &\cdots &|\Delta_0\pi_0\rangle & |\Delta_0\pi_0\pi_0\rangle &\cdots   \\\hline
938	&	0.90        &	-0.42        &	0.10        &	-0.02        &	 	\cdots	&-0.02        &	0.00        &	\cdots \\
1440	&	0.40        &	0.72        &	-0.49        &	0.14        &	 	\cdots	&0.21        &	-0.07        &	\cdots \\
1710	&	0.14        &	0.43        &	0.33        &	-0.21        &	 	\cdots	&-0.74        &	0.28        &	\cdots \\
1975	&	0.07        &	0.29        &	0.60        &	-0.45        &	 	\cdots	&0.57        &	-0.06        &	\cdots \\

\vdots          &\vdots          &\vdots     &\vdots     &\vdots         &\vdots   &\vdots                 &\vdots       &	\ddots     \\
\hline
\end{array}
\end{equation}
and their percentages of each Fock state are
\begin{equation}\label{density_matrix}
\begin{array}{c|ccccccccc}
\hline
E({\rm MeV})    & |p_0\rangle   &|N_0\pi_0\rangle   & |N_0\pi_0\pi_0\rangle &|N_0\pi_0\pi_0\pi_0\rangle &\cdots &|\Delta_0\pi_0\rangle & |\Delta_0\pi_0\pi_0\rangle &\cdots   \\\hline
938	&	0.81        &	0.18        &	0.01        &	0.00        &	 	\cdots	&0.00        &	0.00        &	\cdots \\
1440	&	0.16        &	0.52        &	0.24        &	0.02        &	 	\cdots	&0.04        &	0.00        &	\cdots \\
1710	&	0.02        &	0.19        &	0.11        &	0.04        &	 	\cdots	&0.55        &	0.08        &	\cdots \\
1975	&	0.00        &	0.09        &	0.36        &	0.20        &	 	\cdots	&0.32        &	0.00        &	\cdots \\
\vdots          &\vdots          &\vdots     &\vdots     &\vdots         &\vdots   &\vdots                 &\vdots       &	\ddots     \\
\hline
\end{array}
\end{equation}

Thus, for N(1440), the probability to find it in each Fock state is
\begin{equation}\label{2nd} 
	0.16|p_0\rangle\langle p_0|+0.52|N_0\pi_0\rangle\langle N_0\pi_0|+0.24|N_0\pi_0\pi_0\rangle\langle N_0\pi_0\pi_0|+\cdots+0.04|\Delta_0\pi_0\rangle\langle \Delta_0\pi_0|+\cdots,
\end{equation}
which may be associated with the decay modes in a way like
\begin{equation} 
	\begin{array}{l|c|c}
	\hline
	{\rm Fock\ State}&{\rm Decay\ Mode}&{\rm Final\ State}\\
	\hline
	|N_0\pi_0\rangle	&	N\pi				&	N\pi	\\
	|\Delta_0\pi_0\rangle	&	\Delta\pi			&	N\pi\pi	\\
	|N_0\pi_0\pi_0\rangle	&	N\pi\pi,\ \Delta\pi,\ N\rho	&	N\pi\pi	\\
	|p_0\rangle		&	-				&	-	\\
	\hline
	\end{array}
\end{equation}
Here we need to drop component $|p_0\rangle$ and let Eq. (\ref{2nd}) be normalized as 
\begin{equation} 
	0.63|N_0\pi_0\rangle\langle N_0\pi_0|+0.29|N_0\pi_0\pi_0\rangle\langle N_0\pi_0\pi_0|+\cdots+0.05|\Delta_0\pi_0\rangle\langle \Delta_0\pi_0|+\cdots
\end{equation}
to make an estimation:
\begin{equation}
        \begin{array}{l}
    	{\rm B}\left({\rm N(1440)}\to N\pi\right)          	=0.63,				\\
    	{\rm B}\left({\rm N(1440)}\to N\pi\pi\right) 		=0.29+0.05=0.34.
        \end{array}
\end{equation}
For detailed study of how N(1440) decays, one also needs to know the pion wavefunction. The bare pion $\pi_0$ is not exactly the same as the physical pion $\pi$. Actually, $\pi_0$ here is some like a diquark \cite{diquark2} given its energy $E_{\pi_0}=426$MeV.

If one applies the same argument to N(1710), one will get
\begin{equation} 
        \begin{array}{l}
    	{\rm B}\left({\rm N(1710)}\to N\pi\right)          	=0.19,				\\
    	{\rm B}\left({\rm N(1710)}\to N\pi\pi\right) 		=0.68,				\\
        \end{array}
\end{equation}
which is also in agreement with the experimental observations \cite{PDG}
\begin{equation}
        \begin{array}{l}
    	{\rm B}\left({\rm N(1710)}\to N\pi\right)          	=10\!-\!20\%,				\\
    	{\rm B}\left({\rm N(1710)}\to N\pi\pi\right) 		=40\!-\!90\%.				\\
        \end{array}
\end{equation}
But unlike N(1440), N(1710) also has decay modes like $\Lambda K$. So a detailed study of N(1710) needs one to add in the Fock states that contain strange quarks.

Compare Eq. (\ref{eigv}) with Eq. (\ref{eigv1}), we see that the inclusion of $\Delta_0$ has little effect on the proton and not much effect on N(1440). 
So, if one further adds in heavier Fock states, their effects on the two particles should be even less. 

By Eq. (\ref{density_matrix}), we see that the proton has  81\% of 3-quark Fock state and 19\% of others. Zou \cite{Zoubs_p} has similarly concluded that the probability of multi-quark components in the proton is at least 15\%. By the same equation, we also see that N(1440) is dominated by Fock state $|N_0\pi_0\rangle$ or $|qqqq\bar{q}\rangle$. Jaffe and Wilczeck have similarly suggested in a diquark model \cite{diquark1} that it is a five-quark state $[ud]^2\bar{d}$; Krehl {\it et. al.} have commented \cite{Krehl} that the baryon-meson states plays a role; a recent calculation \cite{Ramalho} of the form factors of $\gamma N\to N(1440)$ suggests that the meson cloud contributions are significant in the region $Q^2<1.5 {\rm\ GeV}^2$. 

\section{Conclusion}
We use the meson cloud model to study together the proton sea quark flavour asymmetry and some properties of N(1440). In the calculation, instead of using the perturbation theory, we construct a Hamiltonian matrix and solve the corresponding eigenvalue equation. The eigenvalues are obtained as the energies of the proton and N(1440), the eigenstates are obtained as their wavefunctions. 
Our study first starts with a simple meson cloud model that contains only bare nucleon and pions. In this case, there are only 3 parameters with which we reproduce the proton sea quark flavour asymmetry and the mass of N(1440).
Then we study with a meson cloud model that also contains $\Delta_0$. In this case, there are 3 more parameters, and we fix them by reproducing the mass of N(1770) and two decay branching ratios: ${\rm B}\left({\rm N(1440)}\to N\pi\right)$ and ${\rm B}\left({\rm N(1440)}\to N\pi\pi\right)$. The inclusion of $\Delta_0$ has not much effect on the wavefunctions of the proton and N(1440). Our study shows that the proton sea quark flavour asymmetry and some properties of N(1440) can be studied in one scheme.

{\bf Acknowledgment} One of the authors is very grateful to Bing-Song Zou and Wei-Zhen Deng for very fruitful discussions. This work was supported by Liaoning Education Office Scientific Research Project(2008288), and by SRF for ROCS, SEM.  The work of B.~Z. was supported by the National Natural Science Foundation of China under Grant No. 10705017.

%
%
%
%
\end{document}